\documentclass[runningheads]{llncs}
\usepackage[normalem]{ulem}
\usepackage{paralist}
\usepackage{censor}
\usepackage{tikz}
\usepackage{hyperref}

\usepackage{enumerate}
\usepackage{booktabs}
\usepackage{subfigure}
\usepackage[breakable]{tcolorbox}
\usepackage{graphicx}

\newif\ifblind
\blindtrue 

\newcommand*{\knowstat}[2]{%
  \medskip

  ---\hfill Knowledge area: \emph{#1}%
  \begin{quote} \emph{#2} \end{quote}	
}

\newcommand{\ListItem}[2]{%
  
  \smallskip
  \begin{tcolorbox}
    \begin{description} \setlength{\partopsep}{0pt} \setlength{\topsep}{0pt}
    \item[#1:] \emph{#2}
    \end{description}
  \end{tcolorbox}
}

\newcommand{\ListSubItem}[2]{%
  \begin{tcolorbox}
    \begin{description} \setlength{\partopsep}{0pt} \setlength{\topsep}{0pt}
    \item[\protect\hspace{1em}#1:] \emph{#2}
    \end{description}
  \end{tcolorbox}
}

\newcommand{\ExplItem}[1]{%
  
  \smallskip
  \begin{quote}
    \noindent \emph{Explanation:} #1
  \end{quote}
}

\begin{document}
\title{Programming for All: \\
  Understanding the Nature of Programs
}

%
%

\author{Andrej Brodnik\inst{1,2}\orcidID{0000-0001-9773-0664} \and
  Andrew       Csizmadia\inst{3}\orcidID{0000-0002-9779-055X}\and
  Gerald       Futschek\inst{4}\orcidID{0000-0001-7255-2531}\and
  Lidija       Kralj\inst{5}\orcidID{0000-0001-5750-6123}\and
  Violetta     Lonati\inst{6}\orcidID{0000-0002-4722-244X}\and
  Peter        Micheuz\inst{7}\orcidID{0000-0002-4722-244X}\and
  Mattia       Monga\inst{7}\orcidID{0000-0003-4852-0067}
}
\authorrunning{A.\ Brodnik et al.}
%
\institute{%
  University of Primorska, Koper, Slovenia      \and
  University of Ljubljana, Slovenia      \and
  Newman University, Birmingham, United Kingdom \and
  TU Wien, Institute of Information Systems Engineering Vienna, Austria \and
  Udruga ``Suradnici u učenju'', Croatia \and
  Università degli Studi di Milano, Milan, Italy
}

\newcommand{\nop}{NoP} 

\maketitle              
\begin{abstract}
	Computer programs are part of our daily life, we use them, we provide them with data, they support our decisions, they help us remember, they control machines, etc. Programs are made by people, but in most cases we are not their authors, so we have to decide if we can trust them. Programs enable computers and computer-controlled machines to behave in a large variety of ways. They bring the intrinsic power of computers to life. Programs have a variety of properties that all citizens must be aware of. Due to the intangible nature of programs, most of these properties are very unusual, but important to understand the digital world. In this position paper, we describe the Nature of Programs in the form of knowledge statements, accompanied by examples from everyday life to clarify their meaning. Everything is formulated in an easily understandable manner and avoids obscure technical language. We suggest that these knowledge statements must be imparted to all teachers and school students.
  A great way to learn and experience the nature of programs is to develop programs yourself.
 
\end{abstract}

\section{Introduction}
\label{sec:intro}


The expression ``Nature of Programs'' (\nop) in the title draws inspiration from  ``Nature of Science'' (NoS)~\cite{nsta}, a similar expression born in the '70s that refers to the fundamental characteristics of science knowledge and scientific inquiry, as derived from how it is produced: a necessary knowledge to make informed decisions with respect to the ever-increasing scientifically-based personal and societal issues. NoS is a significant component of scientific literacy and it is argued that NoS cannot be learned simply by studying science concepts or attending science labs, but it must be addressed explicitly with active reflective practice and discussions among students in their learning contexts.~\cite{Lederman}. This also implies that teachers should have a ``shared accurate view of NoS'' and agree that NoS needs to be taught and assessed explicitly.~\cite{nsta}

Informatics\footnote{In this paper we use the terms `informatics' (preferred) and `computer science' as synonyms.} differs from the natural science in many aspects (at least because its objects are artificial and not natural) and has indeed many peculiarities that ask for a specific approach in education. However, despite the numerous initiatives aiming at popularizing informatics and  introducing it in educational systems, 
there is no clear agreement of what the ``Nature of Computer Science'' is, and the debate on this topic doesn't seem to attract nowadays much interest in our community. Some very significant contributions on this topic appeared in the '80--90s~\cite{shaw1985nature,denning1989computing,turkle1990epistemological} but they need an actualization, if nothing else because they refer to a time where digital devices had a much less relevant role in society than today.
A recent review of the positions on this issue can be found in~\cite{rapaport}.


The centrality of programming in informatics can be found also in most computing education initiatives, which indeed often include some type of programming activity, mainly under the term  \emph{coding}. One can even argue that, for many teachers, computer science is a just a synonym for coding~\cite{cacm-2020}.
Another fundamental component of computing education revolves around the \emph{computational thinking} idea~\cite{wing}. 
Even if there is no clear definition for this expression either, this idea concerns the ability 
to address ``problems in a way that enables us to use a computer and other tools to help solve them''~\cite{ct-flyer}. This includes also the ability to represent, organize, and logically analyze data and to automatize solutions through algorithmic thinking, all skills that have  a fundamental role also in the process of designing programs.


However, the full understanding of what the \nop\ is might not be a natural learning outcome of informatics activities, similarly as it was shown for the NoS~\cite{Lederman}. For instance, practicing coding in a visual programming environment does not imply that students are able to recognize that the programs they write have the same nature as the apps they use on their mobile phones. Similarly, unplugged activities aimed at developing computational thinking skills might be perceived as disconnected from the use of digital devices in everyday life~\cite{benari-unplugged,monga15:issep}.

On the contrary, we claim that understanding the \nop\ is a critical component of the \emph{computing literacy/culture}, and should be one of the main outcome of computing education, alongside with the development of problem solving and computational thinking skills. Moreover we think it should be a core part of any effort to bring digital competences to the masses, like~\cite{DigComp-2.0,DigComp-2.1}, and, even more so, of the initiatives aimed at the inclusion of informatics as a foundational discipline in schools~\cite{cs4all,inf4all}.

In Section~\ref{sec:nop} we discuss the \nop, then in Section~\ref{sec:knowledge} 
we present a series of ``knowledge statements'' that aim at providing a summary about the nature of programs, and are targeted to the general public, starting with teachers and students. Finally, Section~\ref{sec:end} concludes the paper with a summary of our position and hints for future work.

\section{Programming, programs, and the Nature of Programs}
\label{sec:nop}

In fact, the \nop\ makes them distinct from any other human artifacts. Indeed, one can recognize three different facets of programs. 
First, they are \emph{real entities}, in that they affect our everyday lives in the same way as physical or political reality does~\cite{lessig2000code}; 
next, they are \emph{concrete artifacts}, in that they are written, encoded, stored on some digital device;
finally, they are \emph{abstract entities} that manipulate abstract entities (data and data structures). Thus, understanding \nop\ includes grasping the relation between a program and its interpretation by a computing device, being aware of the the duality of instructions and data, accepting the need for and the power of extreme precision and unambiguity, and the unavoidable abstraction involved in computational problems. 

Let's precise better the three facets of programs and why it is important to grasp them all.

Programs are \emph{real entities}: this is probably the most obvious aspect that no one can really ignore. As famously stated by Larry Lessig, nowadays ``code is law''\cite{lessig2000code}, and it is one of the main forces that shapes are daily life. And yet, programs are \emph{designed} by some to be used by many. The goals, the values of the designers are brought into the reality by their translation into algorithms, programs, systems. Indeed, programs are usually created with the purpose of helping in the execution of tasks or the solution of problems (more precisely, they are designed to \emph{automatize} the execution of tasks or the solution of problems).
This is the prevailing facet of programs from the point of view of their users; when using an app on a mobile device or using a piece of software on a personal computer, the users see the program as a useful tool, that has impact on the reality of their job or personal life. This facet is central when talking about digital competences\cite{DigComp-2.0,DigComp-2.1}, where the core is indeed in the ability to use programs for one's own goals. 

Programs are also \emph{concrete artifacts}, in that they are written, encoded, and stored; to use them, one has to have access to them. 
Clearly, the concrete nature of programs is different from the nature of physical artifacts, in that programs are intangible; for instance it's in principle very easy to make copies of programs and modifications can be also cheap or even unconscious (or malicious): we use term \emph{software} for a reason. Thus, programs take up space in the memory of digital devices, they can be deleted, they can be corrupted, and checking their integrity requires special cautions, very different from those that are necessary for checking physical integrity. 
Moreover a program, by its nature, needs an interpreter to be enacted. 
	
Finally, there is the facet of programs that programmers and computer scientists typically deal with and know the most, but the general public often tend to ignore: programs are \emph{abstract entities}. It is indeed very difficult to appreciate the consequences of this without having any acquaintance with the activities involved in programming. Indeed good programs are able to \emph{capture} their users in the abstract world they create, and the relationship with reality becomes hidden enough without an explicit effort of trying to understand \emph{why} something is designed in a given way. Programs are descriptions that must be expressed using the formal rigorous language (the \emph{programming languages}) that the automatic interpret is able to follow. Moreover they manipulate abstract entities, namely data and data structures. Even though these abstract entities do model real entities in general, when modeling them it is necessary to make key choices, simplifying and distorting reality, to make the automatic processing of information possible or feasible.



The process of creating programs is called \emph{programming}.
Unfortunately, for many educators and the general public, programming boils down to coding, \textit{i.e.,} writing instructions in a programming language. However, programming is a much more complex and rich process; 
it consists of \emph{designing} and \emph{developing} a sequence of instructions that a computing device is able to execute, in order to solve automatically a given problem, or perform a given task.
Designing and developing involve (\cite[p.~3]{dagiene-hromkovich-2021}):
`analysis and understanding of problems, identifying and evaluating possible solutions, generating algorithms, implementing solutions in the code of a particular programming language, testing and debugging'.



\section{Knowledge statements}
\label{sec:knowledge}


%
To describe \nop\ in more details, we identify several phases or steps
in program development and its use.
%
%
%
%
%
%
%
%
Reflecting on the development phases of programs and their use, we
created and curated a list of knowledge sentences, which we grouped
into five knowledge areas:
\begin{enumerate}
\item The area \emph{Algorithms} covers the modeling of real problems
  as computational problems, the design and evaluation of their
  computational solutions, and related complexity and computability
  issues.

\item The area \emph{Programs are made of instructions} concerns the
  fact that an algorithm needs to be translated into a programming
  language.  The artifact resulting from this translation is the
  actual program, which is understood and can be executed by the
  automatic interpreter.

\item \emph{Relation between data and program} area addresses how
  programs process abstract data to get output data from input data,
  the relationship between data and the information they represent,
  and ultimately their dual nature of being program and data at the
  same time.

\item In \emph{Programs are running on computers} area, the focus is
  on the distinction between software (programs) and hardware
  (computing devices), and the role of programs in the use of
  computing devices.

\item Finally, \emph{Program execution} area is about the execution of
  programs on a computing device and the effects its execution causes,
  including possibly malicious effects caused by malware or errors.
\end{enumerate}
%
In the rest of this section we
limit ourselves to a sample of statements from different areas, with
examples that exemplify them. The full list of all sentences including
sample skills and attitudes is in the appendix.

\begin{tcolorbox}[breakable]
  \knowstat{Algorithms}{%
    Knowing that to solve a real-world problem on a computer, it has
    to be modeled first in the digital world as a computational
    problem (computational twin).
  }
  \emph{Example.}
  To create a method to find a path through a hedge maze, it is
  advisable to reduce the maze to its essential elements that are
  important for finding a way out. So, the kind of hedge, its height,
  the length and form of the paths, \textit{etc.,} are not relevant. Relevant
  for the model is the structure of the maze, \textit{i.e.,} which walls are
  connected to each other and which part of walls are outside the
  maze. The computational problem is to find a wall that is outside
  the maze.

  \knowstat{Programs are made of instructions}{%
    Knowing that programming languages provide structures that allow
    program instructions to be executed in sequence, repeatedly, or
    only when a certain condition is met.
  }
  \emph{Example.}
  For instance, in a video game a sprite moves around. When it moves,
  it first makes a step with the left foot (one program instruction)
  and then a step with the right foot (the second program
  instruction). The double step represents a sequence of two program
  instructions. Further, when the sprite has to move for 10 double
  steps, it can do this with ten repetitions of the above
  sequence. Moreover during its move, if the sprite bumps into a wall,
  it cannot make a step. To detect the wall, it checks the wall detect
  condition.

  \knowstat{Relation between data and program}{%
    Being aware that the data (e.g. numbers, text, images, sounds) has
    to be properly digitized (digitally encoded), so that it can be
    processed by a program.
  }
  \emph{Example:}
  A picture can be digitized as a grid of color dots. Each color can
  be modeled as a combination of three basic colors and therefore a
  color dot is encoded as a triplet of numbers, indicating the
  brightness of the basic colors. The RGB model uses red, green and
  blue as basic colors.

  \knowstat{Programs are running on computers}{%
    Knowing that computers consist of hardware and software.
  }
  \emph{Example:}
  We want to play a computer game.
  In the first situation we have a gaming device which is just a piece
  of hardware. When we switch it on, it loads and runs a piece of
  software like our gaming application and only then can we play the
  game.
  On the other hand in the second situation we can have a computer
  game (software) on our memory stick. Until the memory stick is
  plugged into the device (piece of hardware), and the device loads
  the computer game from the memory stick and runs it, we can not play
  the game either.
  From the above examples we conclude that we need both hardware and
  software to play the game.
\end{tcolorbox}

Computer programs enable computers and computer-driven devices to
fulfill a wide range of tasks. But everybody should be aware that also
the power of programs is limited. We mention here as examples two
knowledge statements that are related to the limitation of problem
solving by a computer program. The first statement captures the limits
of finding an exact solution to a given problem.
\begin{tcolorbox}[breakable]

  \knowstat{Algorithms}{%
    Knowing that exact solutions to problems may not exist, or may not
    be known, or may require too much time to be computed. They are in
    practice replaced by approximate solutions (this is most likely
    the case with Artificial Intelligence (AI) applications).
  }
  \emph{Example 1.}
  Since a navigation system cannot calculate the exact arrival time in
  advance, because there are many unknown influence factors
  (\textit{e.g.,} changing traffic conditions, congestion), it estimates the
  approximate arrival time.

  \emph{Example 2.}
  Since in some cases it would take too much time for a chess computer
  to compute the best move, it makes an estimation of the best move to
  make.

  \emph{Example 3.}
  In AI and machine learning, deep neural network programs are a
  frequent approach used for sentiment (emotion) recognition from
  images of faces. However, the recognised sentiment is only the best
  guess the algorithm can make.
\end{tcolorbox}

The second statement exemplifies limitation related to the fact that
programs are programmed by humans. Everybody should be aware that
malfunction of computer systems can also be provoked by either the
misbehaviour or computer programmer's mistakes.
\begin{tcolorbox}[breakable]
  \knowstat{Program execution}{%
    Being aware that program execution can result in a computer system
    malfunction, leaking of information or damage; and this may be
    intentional by the design of the programmer (\textit{e.g.,} viruses or
    malware) or unintentional as a consequence of programmer’s
    error. Be aware that programs have to be updated to protect the
    computer system.
  }
  \emph{Example 1.}
  It happens sometimes to us that our text editing program used to
  frequently crash and we lost the result of our work. After the
  program upgrade this never happened again.

  \emph{Example 2.}
  Also it can happen to us that, after opening an email with a
  meaningless attachment, we forget about the message.  After
  a while we get a message from a local pizza store, complaining that
  we are trying to break into their system for placing orders. It
  turned out that the meaningless attachment was a program
  specifically designed to install on our computer a special program
  permitting malicious operators to use our computer in attacking the
  local pizzeria. Because the program was using a flaw in the design
  of the operating system installed on our computer system, the
  problems went away after an update of our computer system.
\end{tcolorbox}

To wrap up, in this section we presented five knowledge areas crucial to
understanding the Nature of Programs and programming. The areas were
illustrated with a few knowledge sentences to give the reader an idea
about them. The list of sentences in each area is longer and covers
all relevant aspects to understand the nature of programs.


\section{Conclusions}
\label{sec:end}

A competent and well-informed citizen of the digital world has to
understand the \nop, which encompasses the many properties of programs
everyone has to be aware of.
%
%
Therefore, also in the K-12 computing education the understanding of
\nop\ must be explicitly addressed together with computational
thinking and problem-solving skills.

In order to elicit the \nop\ and make it accessible to the general public, we wrote a series of knowledge statements describing the properties of programs. Reflecting the development of programs and their use, we grouped the statements into five knowledge areas. 
To illustrate the statements, in this position paper we presented a sample of them through real cases.

We claim that in order to understand the \nop, hence perceiving all the facets that characterize programs, one needs to experience some programming activities first hand. This must include both coding and activities related to the more abstract aspects as algorithms and data representation, as mentioned in Section~\ref{sec:knowledge}. 
Note however that programming activities alone might not be sufficient, and active reflections about the \nop\ should also be fostered in educational contexts.
This clearly brings into question the need
(1) to find 
the appropriate ways to make students appreciate the \nop, and
(2) to properly train teachers so that they acquire  a correct understanding of the \nop\ themselves.



Fully grasping the \nop\ is essential for anyone in order to become digitally conscious, to make sense of the digital world and, ultimately, to act proactively and creatively in it.





%

%
%
\bibliographystyle{splncs04}
\bibliography{p4all}

\appendix
\section{Knowledge areas}
\label{sec:knowledge}

Five knowledge areas are:
\begin{enumerate}
\item The area \emph{Algorithms} covers the modeling of real problems
  as computational problems, the design and evaluation of their
  computational solutions, and related complexity and computability
  issues.

\item The area \emph{Programs are made of instructions} concerns the
  fact that an algorithm needs to be translated into a programming
  language.  The artifact resulting from this translation is the
  actual program, which is understood and can be executed by the
  automatic interpreter.

\item \emph{Relation between data and program} area addresses how
  programs process abstract data to get output data from input data,
  the relationship between data and the information they represent,
  and ultimately their dual nature of being program and data at the
  same time.

\item In \emph{Programs are running on computers} area, the focus is
  on the distinction between software (programs) and hardware
  (computing devices), and the role of programs in the use of
  computing devices.

\item Finally, \emph{Program execution} area is about the execution of
  programs on a computing device and the effects its execution causes,
  including possibly malicious effects caused by malware or errors.
\end{enumerate}

In the rest of the section we discuss each of the knowledge areas in
detail. For each area we give a list of knowledge sentences
with an additional exaplanation where necesary. The lists are
accompanied with lists of skills and attitudes.

\subsection{Algorithms}

\begin{flushright}
  \emph{Knowlede sentences}
\end{flushright}
\ListItem{K1.1}{Knowing that to solve a real-world problem on a
  computer, it has to be modeled first in the digital world as a
  computational problem (computational twin).}
\ExplItem{o create a method to find a path through a hedge maze, it is
  advisable to reduce the maze to its essential elements that are
  important for finding a way out. So, the kind of hedge, its height,
  the length and form of the paths, etc. are not relevant. Relevant
  for the model is the structure of the maze, i.e. which walls are
  connected to each other and which part of walls are outside the
  maze. The computational problem is to find a wall that is outside
  the maze.}

\ListItem{K1.2}{Knowing that an algorithm represents a solution to a
  computational problem.}
\ExplItem{A well known algorithm to find a way through any hedge
  maze, that is enter a hedge maze and exit it at the end, is the wall
  follower algorithm. If one enters a hedge maze then all the walls
  that are connected to each other at the left side of the entrance
  can be walked around by simply following the walls on the left side
  of our walk. As the wall on the left side of the entrance is an
  outside wall, and since all walls are connected, walking around them
  will definitely bring us out again.
 
  This simple approach finds a way through any possible hedge maze, so
  it is an algorithmic solution to a computational problem of maze
  traversal.}

\ListItem{K1.3}{Knowing that various problem-solving strategies are
  used to devise and design algorithms.}
\ExplItem{ When we play cards, we want to have them ordered in our
  hand. One way to order them is to first pick up the smallest card
  and put it on the left, then pick up the second smallest and put it
  next to it and so on. This is an example of a greedy strategy. The
  same strategy works for other completely different problems. For
  example when a cashier wants to give us back change in as few coins
  as possible.

  Other well known problem-solving strategies are divide-and-conquer,
  depth-first-search, breadth-first-search.}

\ListItem{K1.4}{Knowing that an algorithm is presented in a form that
  is language independent but permits a judgment of its correctness
  (that it does what it is supposed to do) and efficiency (how
  efficiently it solves the problem).}
\ExplItem{ We have a shuffled deck of cards and we wonder if the queen
  of spades is in it. We do this by checking each card in the deck in
  order until we find it or we run out of cards. This algorithm can be
  presented e.g. in the form of a flowchart or pseudo-code. These
  representations cannot be executed by a computer, but they should be
  exact enough that one may judge its correctness and efficiency. On
  one hand we can judge that the algorithm will always find the
  searched for card as it checks all cards in the deck. On the other
  hand, to measure the efficiency of the algorithm, we count the
  number of cards we check. The number of cards we check depends on
  how soon we find the queen, but we will never check more cards than
  there are in the deck. Note that if the deck is not shuffled but
  already ordered, we can design a more efficient algorithm to find
  the queen of spades.  }
  
\ListItem{K1.5}{Knowing that an algorithm requires time and space
  (hardware resources) to solve a problem depending on the problem’s
  size.}
\ExplItem{ An algorithm that calculates the sum of an arbitrary
  sequence of values may start with the first one, then add one value
  at a time. The time this algorithm needs depends on the number of
  values in the sequence, while the space, that is memory, it needs
  has to be sufficient to record the sum.  }
  
\ListItem{K1.6}{Knowing that computational problems:}
\ListSubItem{a}{may be solved by different algorithms.}
\ExplItem{ If we are looking for the queen of spades in an ordered
  deck of cards, we can search for it by inspecting one by one each
  card in the deck as explained above. A completely different
  algorithm could first inspect the card in the middle of the deck and
  if it is smaller than the queen of spades it proceeds with a search
  in the upper part of the deck which is just half of the size of the
  original deck. Otherwise, the middle card is smaller than the queen
  of spades, it proceeds in the lower part of the deck. The algorithm
  simply repeatedly applies the same approach until it finds the queen
  or runs out of the deck.}
\ListSubItem{b}{exist that cannot be solved in a reasonable time, even
  for their modest size.}
\ExplItem{
  %
  \begin{figure}[htb]
    \centering
    \includegraphics[width=0.5\textwidth]{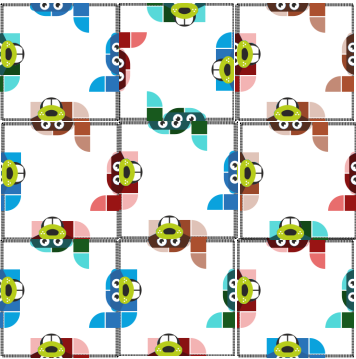}
    \caption{Move pieces to get beavers.}\label{fig-beavers}
  \end{figure}
  A very impressive example is a puzzle like the one shown in Fig.\
  \ref{fig-beavers}. The player gets the image cut into several
  pieces, and she must rebuild these pieces into one image in which
  all beavers fit. In the above picture the colour of the beavers
  (brown, red, blue and green) and the top and the bottom part of the
  head must match. The 3x3 puzzle is not easy to solve for humans, a
  5x5 puzzle is practically impossible. To solve a 7x7 puzzle of this
  kind is even for the fastest available computers out of reach,
  except by chance.  }
\ListSubItem{c}{exist that cannot be solved in principle, no matter
  how fast the machine is or how smart the algorithm designer is.}
\ExplItem{ Peter is a teacher of computer science and he wants to
  check that the programs submitted by the students are
  correct. Therefore he wants to write a program that could check for
  any other program whether it is correct or not
  correct. Unfortunately such a program cannot be written in
  principle.

  There are other computational problems that are also not solvable in
  principle, e.g. automatic checking for any program whether it is
  harmful to our computer (malware), finding all errors in any given
  computer program, automatically writing a program for any given
  computational problem.  }
  
\ListItem{K1.7}{ Knowing that exact solutions to problems may not
  exist, or may not be known, or may require too much time to be
  computed. They are in practice replaced by approximate solutions
  (this is most likely the case with Artificial Intelligence (AI)
  applications).}
\ExplItem{ Since a navigation system cannot calculate the exact
  arrival time in advance, because there are many unknown influence
  factors (e.g. changing traffic conditions), it estimates the
  approximate arrival time.
  
  Since in some cases it would take too much time for a chess computer
  to compute the best move, it makes an estimation of the best move to
  make.
  
  In AI and machine learning, deep neural network programs are a
  frequent approach used for sentiment recognition from images of
  faces. However, the recognised sentiment is only the best guess the
  algorithm can make.
}

\begin{flushright}
  \emph{Skills}
\end{flushright}

\ListItem{S1.1}{Being able to write down instructions to sort a
  deck of cards.}

\ListItem{S1.2}{Given an algorithm that reads up to three numbers,
  makes different calculations and prints the result at the end, being
  able to determine what will be printed when given input numbers are
  entered.}

\ListItem{S1.3}{Given a set of algorithm blocks, being able to combine
  them to solve a more complex problem.}

\ListItem{S1.4}{Given an algorithm with a single loop that sometimes
  runs forever, being able to describe conditions when this happens.}

\ListItem{S1.5}{Having two algorithms that find a card in a sorted
  deck of cards, being able to argue which of them requires more
  steps.}

\begin{flushright}
  \emph{Attitude}
\end{flushright}

\ListItem{A1.1}{Appreciating the benefits of using algorithms in daily
  life.}

\ListItem{A1.2}{Valuing positive and negative aspects of decision
  making algorithms.}

\ListItem{A1.3}{Developing a critical awareness that algorithms are
  the humanly made blueprint of programs.}

\ListItem{A1.4}{Willing to accept an algorithm may not be perfect in
  solving the underlying problem.}

\ListItem{A1.5}{Feeling comfortable with decomposition of tasks.}

\subsection{Programs are made of instructions}

\begin{flushright}
  \emph{Knowlede sentences}
\end{flushright}

\ListItem{K2.1}{Knowing that a computer program is written by humans
  according to strict rules in a programming language.}
\ExplItem{
  When programmers write a program they need to know and stick to the
  rules of the programming language they are using. For instance, to
  display “Hello, Europe” on the screen, the rules in Python require
  us to write:
  \begin{quote} \tt
    print("Hello, Europe")
  \end{quote}
  whereas the rules in PHP require us to write:
  \begin{quote} \tt
    echo "Hello, Europe";    
  \end{quote}
  and the rules in Blockly require to combine the following blocks:
  \begin{center}
    \includegraphics[width=0.5\textwidth]{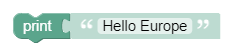}
  \end{center}
  %
  %
  On the contrary, the following writing:
  \begin{quote} \tt
    print (“Hello Europe”]    
  \end{quote}
  does not respect Python rules, because the parentheses are not
  properly matched.
       
  In visual programming environments like Blockly, the shapes of
  blocks make the rules of the language visible and prevent the
  programmer from breaking the rules: if the combination of two blocks
  is not allowed, their shapes prevent them from being connected. For
  instance we cannot directly connect the blocks depicting numbers 12
  and 33 together.  However, the language rules allow you to combine
  them, for example using an operator block:
  \begin{center}
    \includegraphics[width=0.5\textwidth]{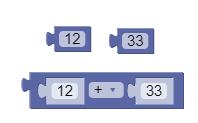}
  \end{center}
  %
}

\ListItem{K2.2}{Knowing that there are different kinds of instructions
  and that there are many programming languages each providing a
  different set of program instructions.}
\ExplItem{There are program instructions that display something on a
  screen, read data from a touchscreen, calculate values, store values
  in memory, etc. Other specific instruction can be provided by
  specialized languages: R language provides program instructions to
  support statistical analysis of data; PHP provides program
  instructions to create interactive webpages; and Scratch provides
  program instructions to display sprites, move them and make them
  interact with each others, as in a video game or a movie.}

\ListItem{K2.3}{Knowing that programming languages provide structures
  that allow program instructions to be executed in sequence,
  repeatedly, or only when a certain condition is met.}
\ExplItem{For instance, in a video game a sprite moves around. When it
  moves, it first makes a step with the left foot (one program
  instruction) and then a step with the right foot (the second program
  instruction). The double step represents a sequence of two program
  instructions. Further, when the sprite has to move for 10 double
  steps, it can do this with ten repetitions of the above
  sequence. Moreover during its move, if the sprite bumps into a wall,
  it cannot make a step. To detect the wall, it checks the wall detect
  condition.}

\ListItem{K2.4}{Knowing that a programmer has to be aware of what
  program instructions do, but not necessarily how they do it.}
\ExplItem{If a Python programmer wants to print text \texttt{“Hello
    Europe”} on the screen, she uses the program instruction
  \texttt{print(“Hello Europe”)}. After this, the programmer needs not
  to worry about what needs to happen in order for the text to appear
  on the screen.}

\ListItem{K2.5}{Knowing that new instructions can be created from
  existing instructions to perform a specific task.}
\ExplItem{In Blockly or in many other programming languages, we can
  create a new program instruction that draws a square. From now on we
  can use the new program instruction as any other program instruction
  in our program. A new program instruction created by a programmer is
  called differently in different programming languages: function,
  subroutine, macro, procedure etc.

  Most programming languages permit created program instructions to be
  packed into libraries which can be shared among programmers and
  provide ready-made program instructions.}

\begin{flushright}
  \emph{Skills}
\end{flushright}

\ListItem{S2.1}{Being able to make changes in a formal description of
  a procedure.}
\ExplItem{Being able to modify or complete a simple program written in
  a visual programming language.}

\ListItem{S2.2}{Giving a procedure described in a formal way, being
  able to recognize which kind of information it processes.}
\ExplItem{Given a simple program written in a known programming
  language, being able to identify if it processes numerical data,
  images, point coordinates, or something else.}

\ListItem{S2.3}{Giving a procedure described in a formal way, being
  able to recognize the flow of information, that is how information
  is processed.}
\ExplItem{Given a simple program written in a known programming
  language and a given input, being able to find out what the output
  will be.}

\ListItem{S2.4}{Being able to detect possible problems in a formal
  description of a procedure.}
\ExplItem{Being able to reproduce the error of a program and perform
  simple debugging.  Being able to identify and describe the
  conditions under which the program written in a known programming
  language does not behave as desired.}

\begin{flushright}
  \emph{Attitude}
\end{flushright}

\ListItem{A2.1}{Developing the awareness that the responsibility of
  software failures should be attributed to the software developers.}

\ListItem{A2.2}{Not being afraid to try to understand the source code
  of a program.}

\ListItem{A2.3}{Discussing and discovering if two different sequences
  of instructions give the same result.}

\ListItem{A2.4}{Appreciating the possibility and value of creating new
  instructions.}
\ExplItem{New instructions can for example be functions.}

\ListItem{A2.5}{Developing an appreciation for expressing actions and
  conditions precisely and unambiguously, even in natural language
  descriptions.}

\subsection{Relation between data and program}

\begin{flushright}
  \emph{Knowlede sentences}
\end{flushright}

\ListItem{K3.1}{Knowing that programs produce results depending on
  input data, and that different input data usually yields different
  output data.}
\ExplItem{In a navigation system the input is a destination location
  and the output is a route to it. Different input destinations yield
  different output routes.}

\ListItem{K3.2}{Being aware that the data (e.g. numbers, text, images,
  sounds) has to be properly digitized (digitally encoded), so that it
  can be processed by a program.}
\ExplItem{A picture can be digitized as a grid of color dots. Each
  color can be modeled as a combination of three basic colors and
  therefore a color dot is encoded as a triplet of numbers, indicating
  the brightness of the basic colors. The RGB model uses red, green
  and blue as basic colors.}

\ListItem{K3.3}{Being aware that there is a difference between data
  and information, in that data has no intrinsic meaning and gives
  information only when interpreted by humans in the context of the
  problem being solved.}
\ExplItem{Data “42” can be interpreted as a temperature if it occurs
  in a sentence like “Today is 42 C degree.” or as an age if it occurs
  in a sentence like “I’m 42 years old.”, or as the answer to the
  Ultimate Question of Life, The Universe, and Everything, if it is
  said by Douglas Adams.}

\ListItem{K3.4}{Knowing that input data models information that is
  relevant to the problem statement, and output data provide
  information relevant to its solution.}
\ExplItem{In the context of a translation program, the input word (or
  sentence) models the meaning that the word (resp., sentence) has in
  the source language, and the output data models the meaning of the
  word (resp., sentence) in the destination language.}

\ListItem{K3.5}{Knowing that a program itself is data.}
\ExplItem{When installing an application, we download data that is
  later executedby the machine as a program.} 

\begin{flushright}
  \emph{Skills}
\end{flushright}

\ListItem{S3.1}{Being able to identify input and output data in some
  simple programs.}

\ListItem{S3.2}{Being able to recognize the difference between “batch”
  programs in which the input is given once for all, and then output
  is produced and “interactive” programs.}

\ListItem{S3.3}{Being able to find information about digital encoding
  of specific data.}

\ListItem{S3.4}{Being able to determine the amount of data to be
  downloaded to install a program.}

\begin{flushright}
  \emph{Attitude}
\end{flushright}

\ListItem{A3.1}{Inclined to critically evaluate the choice of programs
  by the demand and supply of data.}

\ListItem{A3.2}{Discussing the benefits of different choices for
  digital encoding of data.}

\ListItem{A3.3}{Arguing about the amount of data that has to be
  downloaded to install a program by backing up one’s claim with
  evidence.}

\ListItem{A3.4}{Appreciating the importance of context that gives
  meaning to data and computation.}

\subsection{Programs are running on computers}

\begin{flushright}
  \emph{Knowlede sentences}
\end{flushright}

\ListItem{K4.1}{Being aware that many technological devices contain
  computers which control them.}
\ExplItem{For example, a greenhouse is digitally controlled by a
  variety of programmed devices: A smart temperature sensor runs a
  program that checks whether the temperature is above or below the
  target range for a particular time of day and then triggers the
  heating or cooling system when needed. Further, a security camera
  uses a machine learning algorithm to detect non-plant motion and
  sends a video sequence to a computer. Both, the data from the
  temperature sensor and the video sequences are stored on a desktop
  computer which moreover generates a graph of the temperature. All
  devices are connected to a (wireless) router which uses networking
  software to enable the transmission of data.}

\ListItem{K4.2}{Knowing that computers consist of hardware and
  software.}
\ExplItem{We want to play a computer game. In the first situation we
  have a gaming device which is just a piece of hardware. When we
  switch it on, it loads and runs a piece of software like our gaming
  application and only then can we play the game.

  On the other hand we can have a computer game (software) on our
  memory stick. Until the memory stick is plugged into the device
  (piece of hardware), and the device loads the computer game from the
  memory stick and runs it, we can not play the game either.

  From the above examples we conclude that we need both hardware and
  software to play the game.}

\ListItem{K4.3}{Knowing that an operating system is a special type of
  software that enables a user to run application software on a
  computing device.}
\ExplItem{We are using our smart phone to attend a video conference call using
  the phone’s camera, speaker and microphone. In the middle of the
  video conference call comes a phone call and a phone application is
  started which needs to first use the speaker to ring and then, if we
  decide to take the call, now let the phone application also use the
  speaker and the microphone. All the coordination work needed for
  this is done unseen for the user by the operating system and the
  user is barely aware of it. In this way the operating system permits
  several applications to run simultaneously on the device.

  An operating system manages input devices such as a keyboard and
  mouse, output devices such as monitors, speakers and printers and
  storage devices such as internal and external memory. Moreover, it
  also manages network connections. Finally, to interact with the
  operating system the user can use either a graphical user interface
  (GUI) or command line interface.}

\ListItem{K4.4}{Recognizing that computer systems which have the same
  or similar hardware may have different operating systems with
  different (graphical) user interfaces installed on them.}
\ExplItem{When we buy a new computer, we can install either Linux
  operating system, Windows operating system, FreeBSD operating system
  or some other operating system. This is true even for the Apple
  computers and some smartphones. Moreover, when we install the
  operating system, we can also use different graphical interfaces on
  the same operating system, which is most common on the Linux
  operating system.}

\begin{flushright}
  \emph{Skills}
\end{flushright}

\ListItem{S4.1}{Being able to identify which program is responsible
  for the overload of a computer system and stop the program.}
\ExplItem{Being able to launch the task manager utility to get
  information about the resources used by programs that are running on
  your personal computer.}

\ListItem{S4.2}{Being able to install, update and uninstall different
  software on a computer system.}
\ExplItem{Being able to upgrade the version of the operating system
  on a Smart TV.  Being able to uninstall an app from a smartphone.}

\ListItem{S4.3}{Being able to recognize the same simple functionality
  under different operating systems.}
\ExplItem{Being able to list the files in Desktop on iOs and Windows
  MS.}

\begin{flushright}
  \emph{Attitude}
\end{flushright}

\ListItem{A4.1}{Developing a positive attitude and self-confidence in
  the use of programs.}

\ListItem{A4.2}{Being open to new programs, self-confident in
  exploring their functionalities, respecting the experiences and
  opinions of others when selecting new programs.}

\ListItem{A4.3}{Evaluating new program tools before using them and
  confidently switching to them if favourable assessment.}

\ListItem{A4.4}{Helping and guiding others in selecting and using new
  programs.}

\subsection{Program execution}

\begin{flushright}
  \emph{Knowlede sentences}
\end{flushright}

\ListItem{K5.1}{Knowing that a computer is able to automatically
  interpret and execute instructions.}
\ExplItem{The program is a sequence of instructions. The instructions
  are executed one after the other by the computer. In fact the
  computer consists of several parts and one of them is the central
  processing unit (CPU) which interprets and executes instructions
  automatically one at a time.}

\ListItem{K5.2}{Knowing that for a program to be executed by a
  computing device it may be necessary to translate it from the
  programming language it is written in, into a language understood by
  a computing device, and this is done by specific programs.}
\ExplItem{We can run an App that we have downloaded to our mobile
  phone, but we can neither inspect its code nor modify it. In order
  to modify the App, one needs to have access to its source code. In
  order to update an App, its developers modify its source code, then
  use a specific program (compiler) to translate the source code into
  a language understood by mobile phones, and finally release the
  resulting updated version of the App.}

\ListItem{K5.3}{Knowing that execution of programs changes data stored
  in the computer system.}
\ExplItem{Each time you enter or update your emails through the
  software, the program checks for new emails and downloads them if
  there are any new ones, resulting in a change in the data stored
  about emails on your computer.}

\ListItem{K5.4}{Being aware that program execution can result in a
  computer system malfunction, leaking of information or damage; and
  this may be intentional by the design of the programmer
  (e.g. viruses or malware) or unintentional as a consequence of
  programmer’s error. Be aware that programs have to be updated to
  protect the computer system.}
\ExplItem{It happens sometimes to us that our program for text editing
  used to crash frequently and we lost the result of our work. After
  the program upgrade this never happened again.

  Also it can happen to us that after opening an email with a
  meaningless attachment and then we forget about the message. After a
  while we get a message from a local pizza store, complaining that we
  are trying to break into their system for placing orders. It turned
  out that the meaningless attachment was a program specifically
  designed to install on our computer a special program permitting
  malicious operators to use our computer in attacking the local
  pizzeria. Because the program was using a flaw in the design of the
  operating system, the problems went away after an update of our
  computer system.}

\ListItem{K5.5}{Be aware that during the execution of a program the
  data is stored in either a permanent or a volatile way and the one
  stored in the latter is lost when the application crashes or the
  power is cut off.}
\ExplItem{If your computer suddenly crashes (stops working) while you
  are writing a document, for example in a word processor, there is a
  chance that the latest changes to your document are lost because it
  was not stored permanently. Similarly, your e-mail message could be
  lost if the mail program stops working while you are writing
  it. This could happen with almost any program or computing device so
  have in mind to check that it is automatically stored on the device
  or in the cloud frequently enough or you have to store it explicitly
  yourself.}

\ListItem{K5.6}{Being aware that there are programs (applications)
  with a graphical user interface that are used by users to accomplish
  a certain task, and other programs without a user interface that are
  started automatically, and are necessary to make the computer useful
  (such as operating systems, system support programs, programs on
  embedded systems).}
\ExplItem{The software we use from time to time needs to be
  updated. However, we do not know when the update will occur. For
  this reason, a special program starts automatically without an
  explicit user command, and periodically checks the availability of
  an update.

  While we are browsing the Internet an antivirus program may
  automatically check for unexpected attempts to access our
  computer. Similarly, when we want to communicate on social media
  platforms, an integrated grammar checking program may automatically
  correct our posts. Finally, when the posting is ready to be
  published and we press the share button, we need not to care about
  all the details involved in transferring bits on the network
  appliance, as the operating system automatically arranges both
  hardware and software resources to transfer and publish the
  content. The operating system, automatic grammar checkers and virus
  checkers often work without interacting with the user.}

\begin{flushright}
  \emph{Skills}
\end{flushright}

\ListItem{S5.1}{Being able to start execution of a program on a
  personal computer or a smartphone.}

\ListItem{S5.2}{Being able to stop execution of a program on a
  personal computer or a smartphone.}
\ExplItem{Especially when the program cannot be controlled within the
  user interface provided by the program, e.g. endless loop. }

\ListItem{S5.3}{Being able to describe how a used piece of software
  saves user input to be retrieved in the future.}
\ExplItem{For example, saved automatically/by explicit saving by user,
  locally/remotely.}

\ListItem{S5.4}{Being able to describe the context under which an app
  is malfunctioning.}

\begin{flushright}
  \emph{Attitude}
\end{flushright}

\ListItem{A5.1}{Appreciating the benefit of computers automatically
  interpreting and executing instructions.}

\ListItem{A5.2}{Developing a critical attitude to periodically
  checking for a new version of the program.}

\ListItem{A5.3}{Developing cautiousness about downloaded programs.}

\ListItem{A5.4}{Trying to find out why a program is not acting as
  expected and persistently trying to find a solution, even if
  attempts were less successful.}


\end{document}